\documentclass[usenatbib]{basi}
\usepackage[T1]{fontenc}
\usepackage[british]{babel}
\usepackage[varg]{txfonts}
\usepackage{color}
\begin{document}

\title[On the possibilities of {mass loss} from an advection accretion disk]
{On the possibilities of {mass loss} from an advective accretion disc around stationary
black holes}

\author[S.\ Das et al.]{Santabrata
    Das$^1$\thanks{email: \texttt{sbdas@iitg.ernet.in}},
    Indranil Chattopadhyay$^2$, Anuj Nandi$^3$ and Biplob Sarkar$^1$\\
    $^1$ Department of Physics, IIT Guwahati, Guwahati 781039, Assam\\
    $^2$ ARIES, Manora Peak, Nainital 2630 02, Uttarakhand, India\\
    $^3$Space Astronomy Group, SSIF/ISITE Campus, ISRO Satellite Centre, \\
        Outer Ring Road, Marathahalli, Bengaluru 560 037, India}

\pubyear{2014}
\volume{42}

\date{Received 2014 January 21; accepted 2014 May 15}

\maketitle

\label{firstpage}

\begin{abstract}
We study the coupled disc-jet system around the black hole where the
outflow solutions are obtained in terms of the inflow parameters. We observe
that an advective accretion disc can eject outflows/jets for wide range
of viscosity parameter. However, such possibility is reduced if the 
cooling is active as the energy dissipative process inside the disc.
For mass outflow, we obtain the parameter space spanned by the inflow
angular momentum and the viscosity in terms of cooling and
quantify the limits of viscosity parameter.
\end{abstract}


\section{Introduction}\label{sec:intro}
Outflows/jets are commonly observed from black hole systems. They seem to
originate from the accreting matter as black hole does not have any 
intrinsic atmosphere. In rotating accreting matter, centrifugal
force at the vicinity of the black hole acts as a barrier to the
faster supersonic matter following it and eventually
slows it down. If the barrier is strong enough
then it triggers the formation of
shock waves \citep{c96,dcnc01,cd04,d07,c08,cc11}.
This centrifugally supported post-shock matter is hot and
piles up in the form of a torus around the black hole
called CENtrifugal pressure supported BOundary Layer (CENBOL) 
{\citep{c99}}.
In the post-shock region, accreting matter is compressed and becomes
hot. Consequently, excess thermal gradient force develops in the
post-shock disc, which ultimately deflects a part of the accreting
matter to form bidirectional outflows/jets. 
This was shown via Lagrangian, as well as, Eulerian numerical simulation codes
\citep{mlc94,mrc96,lrc11}. Theoretically, \citet{c99} computed the mass outflow rate for
very simplistic flow configurations, in terms of inflow parameters. In the same spirit, but
considering rotating accretion flow,
\citet{dcnc01} computed the mass outflow rates from input parameters
of inviscid flow where the accretion is rotating but jet was assumed to be radial.
Later, \citet{cd07,dc08} calculated mass loss from  
dissipative accretion disc considering rotating outflow. Subsequently, 
mass outflows have been computed for variety of shocks \citep{bdl08,dbl09},
various types of acceleration mechanisms \citep{kc13,kcm14}, and even for variable
adiabatic index \citep{kscc13}, using an equation of state of the flow designed to handle
multi-species fluid \citep{cr09}. 
Recently, \citet{dcnm14} showed that a part of the inflowing matter is
periodically ejected from the disc when the viscosity parameter is chosen beyond its
critical values.
Already, \citet{cd07} and \citet{dc08} studied the outflow properties in
terms of the disc viscosity and cooling processes. The outflow rate 
computed from the appropriate
choice of the inflow parameters is consistent with the jet kinematic
luminosity estimated from the radio luminosity of the jets \citep{hs03}.
Here, we mean outflows as jets and not winds.
Jets differ from winds as they are usually fast, supersonic and collimated
outflows. It is well known that the increase of accretion rate demonstrates
the transition of the black hole sources from hard to softer spectral states 
\citep{ct95,c97} which suggests an
underlying connection between outflow properties and the spectral states. 
The outflow rate also depends on viscosity, which indicates that there
exists a range of viscosity that allows mass loss from the disc for a
given accretion rate. Such critical limits of the viscosity is not
studied so far and for the first time, we identify the range of viscosity 
for the formation of mass outflow from the accretion disc .

In the next Section, we present the basic assumptions and governing
equations. In Section 3, we discuss the results and finally present
the conclusions.  

\section{Assumptions and model equations}

We consider a steady, viscous, axisymmetric accretion 
flow on to a Schwarzschild black hole. We adopt 
\citet{pw80} potential to describe space-time geometry around the 
black hole. Also, the flow is assumed to be in hydrostatic equilibrium in the 
vertical direction \citep{matetal84}. In our model, 
jets are tenuous, and have negligible shear and therefore can be assumed to be
inviscid \citep{cd07} and 
have low angular momentum compared to the outer edge of the disc.
In this work, we use $2G=M_{\rm BH}=c=1$ unit system,  where
$G$, $M_{\rm BH}$, and $c$ are the gravitational constant, the
mass of the black hole, and the velocity of light, respectively.

\subsection{Equations for accretion}
The dimensionless governing equations for accretion are \citep{c96},

\noindent (a) the radial momentum equation:
$$
u \frac {du}{dx}+\frac {1}{\rho}\frac {dP}{dx}
-\frac {\lambda^2(x)}{x^3}+\frac {1}{2(x-1)^2}=0,
\eqno(1a)
$$
\noindent (b) the mass conservation equation:
$$
\dot M = 2 \pi \Sigma u x,
\eqno(1b)
$$
\noindent (c) the angular momentum conservation equation:
$$
u \frac {d\lambda(x)}{dx}+\frac{1}{\Sigma x}
\frac {d}{dx}\left( x^2 W_{x\phi}\right)=0,
\eqno(1c)
$$
and (d) the entropy equation:
$$
u T \frac {ds}{dx}= Q^{+} - Q^{-} 
\eqno(1d)
$$
where, $x$, $u$, $\rho$, $P$ and $\lambda(x)$ are the radial distance, 
radial velocity, density, isotropic pressure and specific angular
momentum of the flow, respectively. Here, $\Sigma$ is vertically
integrated density, 
$s$ is the specific entropy of the flow, and $T$ is the
local temperature. The viscous stress is represented by $W_{x\phi}$
($=- \alpha \Pi$, $\Pi = [W+\Sigma u^2] $ is the vertically integrated
total pressure, $\alpha$ is the viscosity parameter, $W$ is vertically
integrated thermal pressure).
The gain of heat by the flow are given by \citep{c96,dc08},
$$
Q^{+} = -\frac{{\alpha}}{\gamma} x(ga^2+\gamma u^2)\frac{d\Omega}{dx},
\eqno(2a)
$$
where, $\gamma$ is the adiabatic index, $g = I_{n+1}/I_n$, $n=1/(\gamma -1)$,
$I_n=(2^n n!)^2/(2n+1)!$ \citep{matetal84}. In this work, Bremsstrahlung
cooling process is ignored as it is inefficient and only synchrotron cooling
is considered which is given by \citep{dc08},
$$
Q^{-} = \frac{Sa^5}{ux^{3/2}(x-1)},
\eqno(2b)
$$
where,
$$
S = \frac{32\eta\mu^2 e^4 \times 1.44\times10^{17}}{3\sqrt{2}m^3_e\gamma^{5/2}}
\frac{{\dot m}_i}{2GM_{\odot}c^3}.
\eqno(2c)
$$
Here, $e$ is the electron charge, $m_e$ is electron mass, $M_{\odot}$ is
solar mass, and $\mu = 0.5$ for fully ionized plasma. The accretion rate
${\dot m}_i$ is measured in units of Eddington accretion rate adopting 
$10 M_{\odot}$ black hole. We use $\eta~(\le 1)$ which is
the ratio between the magnetic pressure to the gas pressure to estimate
the magnetic field for synchrotron cooling. 
In this work,
$\gamma=4/3$ and $\eta = 0.1$ are used throughout the paper.

\subsection{Equations of motion for outflows}
The conserved energy equation for jet is given by \citep{cd07},

$$
{\mathcal E}_j= \frac{1}{2}v^2_j+na^2_j +\frac {\lambda^2_j}{2x^2_j}
-\frac{1}{2(r_j-1)},
\eqno{(3a)}
$$
where, ${\mathcal E}_j$ and $\lambda_j$
are the specific energy and the angular momentum of the jet, respectively.
The integrated continuity equation for jet is,
$$
{\dot M}_{\rm out}=\rho_j v_j {\mathcal A},
\eqno{(3b)}
$$
where, ${\dot M}_{\rm out}$ is the outflow rate, $\rho_j$ is the jet density,
$v_j$ is the jet velocity and ${\mathcal A}$ denote the area function of the
jet \citep{cd07}.

\section{Method}

Since a part of the accreting matter is deflected at the CENBOL
surface to form bidirectional outflows \citep{dcn13,dcnm14},
therefore, our focus would be on such 
accretion solutions that contain stationary shocks. For shock, accreting
flow must possess two saddle type sonic points, namely inner sonic point
($x_{ci}$) and outer sonic point ($x_{co}$). Interestingly, the range of 
$x_{ci}$ is usually restricted within $2-4~r_g$ ($r_g \equiv$ Schwarzschild
radius) \citep{cd04,d07} and we choose $x_{ci}$ as one of the input
parameters to obtain
the accretion solution that includes shock wave. Other input parameters
are angular momentum $\lambda_i$ at $x_{ci}$, viscosity $(\alpha)$ and
accretion rate ${\dot m}_i$, respectively. We calculate the
accretion solution by integrating the governing equations (Eqs. 1a-d)
once from $x_{ci}$ inwards and then outwards \citep{cd04}. The condition
for steady shock requires conservation of energy flux, mass flux,
and momentum flux across the shock front \citep{ll59} and in
presence of mass loss, conservation of mass flux is taken care
considering mass outflow rate $R_{\dot m}$ defined as the ratio
between the mass flux of the outflow (${\dot {M}}_{\rm out}$)
and the pre-shock accretion rate (${\dot {M}}_{-}$) \citep{cd07} which
are given by,

$$
{\mathcal E}_{+} = {\mathcal E}_{-};~
{\dot{M}}_{+} ={\dot {M}}_{-}-{\dot {M}}_{\rm out}
={\dot {M}}_{-}(1-R_{\dot m}); ~
{\Pi}_{+} = {\Pi}_{-},
\eqno{(4)}
$$
where, subscripts ``$-$'' and ``$+$'' refer
to the quantities before and after the shock, respectively.

We consider that jet is launched with the same energy, angular
momentum and density of the post-shock flow and the corresponding
mass outflow rate is given by \citep{cd07,dc08},
$$
R_{\dot m} =\frac{{\dot M}_{\rm out}}{{\dot M}_-}
=\frac{Rv_{j}(x_s) {\mathcal A}(x_s)}
{4 \pi \sqrt{\frac{2}{\gamma}}x^{3/2}_s (x_s-1)a_+u_{-}},
\eqno{(5)}
$$
where, $x_s$ is the shock location, $R~(={\Sigma_+}/{\Sigma_-})$
is the compression ratio and $a_{+}$ is the post-shock sound
speed. We simultaneously solve the accretion-ejection equations
to obtain shocks in presence of mass loss in the following way.
We first find the virtual shock location $x^{\prime}_{s}$ without
considering mass loss. We assign shock energy ${\mathcal E}(x^{\prime}_{s})$
and angular momentum $\lambda(x_{s}^{\prime})$ as the jet energy ${\mathcal E}_j$
and $\lambda_j$ and solve the jet equations to obtain corresponding
$R_{\dot m}$. Now we use $R_{\dot m}$ in shock condition to find
the new shock location. We continue the iteration to converge for
actual shock location and the corresponding $R_{\dot m}$ is the
relative mass outflow rate. Here, the jet is assumed to have the same
energy as the post shock disc. Since these bidirectional outflows are
launched from the post-shock disc, therefore, the flow parameters at the
jet base should be same as that of the disc from where it is being launched.
For simplicity, we consider the immediate post-shock Bernoulli parameter
to be that of the jet.

\section{Results and discussion}

\begin{figure}
\centerline{\includegraphics[width=8cm]{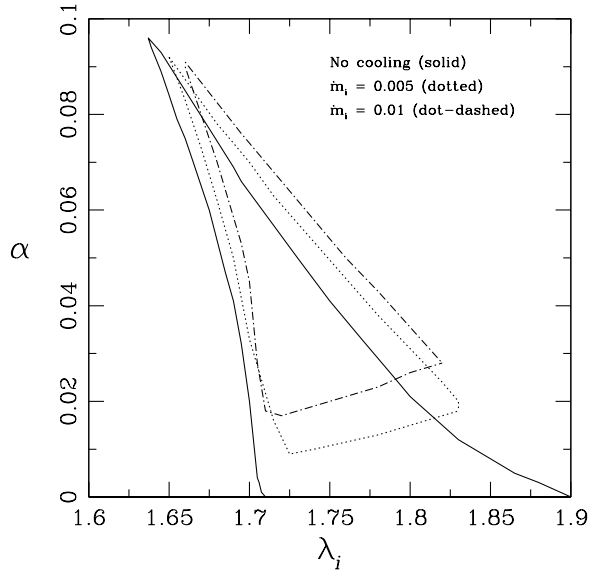}}
\caption{Parameter space spanned by the angular momentum
$(\lambda_{i})$ at the inner sonic point and the viscosity. Different
boundaries separate the parameter space for mass loss from the disc.
Solid boundary is for solutions having no cooling and the boundaries drawn 
with dotted and dot-dashed curves are for various accretion rates marked in
the panel. See text for details.}
\label{f:one}
\end{figure}

In this work, we intend to find the range of viscosity that admits mass
loss from an advective accretion disc around the black holes. To achieve
our goal, we supply the inflow parameters, namely, inner sonic points
($x_{ci}$), angular momentum $(\lambda_{i})$ and viscosity ($\alpha$),
respectively and look for global inflow-outflow solutions. Of the three
parameters, we fixed angular momentum $(\lambda_{i})$ and vary
inner sonic points ($x_{ci}$) and viscosity ($\alpha$) for all possible
range. Essentially, we identify the region of the parameter space
spanned by the angular momentum $(\lambda_{i})$ and viscosity ($\alpha$)
for mass loss which is shown in Fig. 1. It is indeed fascinating to
see that mass loss is possible for a wide range of viscosity
which again strongly depends on the angular momentum of the accreting
matter at the inner part of the disc. We classify the parameter space
as function of accretion rate ${\dot m}_i$ 
and separate it with the solid, dotted and dot-dashed
boundaries. Solid boundary represents the cooling free solutions that
are independent of ${\dot m}_i$.
The dotted and dot-dashed boundaries denote the solutions that correspond to
$ {\dot m}_i= 0.005$ and ${\dot m}_i = 0.01$, respectively.
As the cooling is increased, the size of the post-shock region shrinks
\citep{d07} and the possibility of mass loss is reduced 
\citep{c99,dcnc01} which is consistent with the conclusions of
\citet{dc08} and \citet{ggc12}.
This result has a profound implication as
\citet{dc08} pointed out that with the appropriate choice of the
inflow parameters, the present disc-jet model successfully explains the
jet luminosity at least for two objects, namely, M87 and
Sgr~${\rm A}^{*}$. \citet{dc08} estimated the mass outflow rate of M87 jets as
$0.009 M_{\odot}$ yr$^{-1}$ which gives a kinematic luminosity of
$\sim 10^{43}$ergs s$^{-1}$ \citep{rmfhc96} considering few percent efficiency, while
for Sgr~${\rm A}^{*}$ the estimated mass loss was $9\times 10^{-8} M_{\odot}$ yr$^{-1}$
that also satisfies the kinematic luminosity of $\sim 10^{38}$ergs s$^{-1}$
accepted in the literature \citep{y00,ymf02}.
In the similar way, with this methodology one can estimate the jet power in hard states
for the galactic stellar mass black hole sources as well.

In this work, we choose a set of $\gamma$ and $\eta$ values to 
represent our results. We examined the dependency of mass loss
parameter space on $\gamma$ and observed only quantitative variation
in a way that the parameter space is shifted towards the lower
viscosity and angular momentum side for $\gamma > 4/3$.
As discussed in Section 2.1, the limit is $\eta \le 1$ and we use
$\eta = 0.1$. For higher $\eta$, the effect of radiative loss would be
more for the same accretion rate (Eq. 2c) and thereby reduce the thermal
driving of the jets. 
Eventually, the parameter space for mass loss would further be
reduced with the increase of $\eta$ values.

Recently \citet{nrs13} reported that several GBH sources
exhibit hard to hard-intermediate spectral state transition
along with the increase of QPO frequency and radio
and X-ray fluxes.
However, during the transition from hard-intermediate
to soft-intermediate state, the QPO disappears
followed by the transient radio flare. This perhaps indicates
that the presence of weaker/mildly relativistic, continuous jet
is related to a component of the disc which is responsible for
QPO. And the disappearance of QPO followed by strong radio flare
shows that it is the same component of the disc which gets disrupted and
expelled as strong, relativistic ejections during the
hard-intermediate
to soft-intermediate state transition.
This observation confirms the presence of shock in accretion disc
and also identifies the  post-shock disc (CENBOL) as the seat of
hot electrons/Comptonizing cloud whose dynamics is responsible
for jet generation and QPOs. 
With these insights, it would be intriguing to interpret the relations
among the jet evolution, spectral states and QPOs in greater details.
Indeed, some of the seminal works has already been
initiated \citep{cm00,gc13}.
Our present effort which is the
identification of ranges
of viscosity parameter and accretion rate 
resulting steady jets, would be an insight 
for impending numerical simulations to study variabilities
of black hole accretion-ejection system.
Overall, one can 
limit the value of viscosity parameter while explaining
the jet power for numerous objects which we intend to explore elsewhere.

\section*{Acknowledgements}

AN acknowledges Dr. Anil Agarwal, GD, SAG, Mr. Vasantha E. DD, CDA and Dr.
S. K. Shivakumar, Director, ISAC for continuous support to carry out this
research. The authors also acknowledge the anonymous referee for
fruitful suggestions to improve the quality of the paper.

\label{lastpage}
\end{document}